# FAKE NEWS DETECTION TOOLS AND METHODS – A REVIEW

Sakshini Hangloo[1], Bhavna Arora[2]
Ph.D Scholar[1], Assistant Professor[2]
Email: sakshini.hangloo@gmail.com[1], bhavna.aroramakin@gmail.com[2]
Department of Computer Science & Information Technology
Central University of Jammu, Bagla (Rahya Suchani), District-Samba,
Pin Code 181143, Jammu, J&K, India

**ABSTRACT**

*In the past decade, the social networks platforms and micro-blogging sites such as Facebook, Twitter, Instagram and Sina Weibo have become an integral part of our day-to-day activities and is widely used all over the world by billions of users to share their views and circulate information in the form of messages, pictures, and videos. These are even used by government agencies to spread important information through their verified Facebook accounts and official Twitter handles, as it can reach a huge population within a limited time window. However, many deceptive activities like propaganda and rumour can mislead users on a daily basis. In this COVID times the fake news and rumours are very prevalent and are shared in a huge number which has created chaos in this tough time. And hence, the need of Fake News Detection it the present scenario is inevitable. In this paper, we survey the recent literature about different approaches to detect fake news over the Internet. In particular, we firstly discuss about fake news and the various terms related to it that have been considered in the literature. Secondly, we highlight the various publicly available datasets and various online tools that are available and cam debunk Fake News in real time. Thirdly, we describe fake news detection methods based on two broader areas i.e., it's content and the social context. Finally, we provide a comparison of various techniques that are used to debunk fake news.*

## 1. INTRODUCTION

Social media has become an important information platform from where people gain and share information. Its increasing popularity has also enabled the wide dissemination of misinformation causing significant negative effects on society. Therefore, to maintain social harmony it is highly crucial to detect fake news on these platforms and also regulate these to ensure that the users receive genuine information. The social media users are mostly naïve, and get affected by the false information circulated on these platforms. They may unintentionally share the fake content and support them by commenting on the fake news. Some political experts believe that the victory of Donald Trump in 2016 U.S. presidential election was an outcome of propaganda and rumors[1]. Since last two years also the social media networks in different countries are continuously being flooded with all kinds of fake news and rumors on COVID-19. Today we are living in the information age, where the users are a content generator but most of the information generated lack of trust and verification because of the lack of debunking tools. The fake news detection on social media platforms have become an emerging research topic and is gaining attention of researchers all over the world. Many technical giants also like Google and Facebook are looking for solutions to debunk online fake information. In 2016, Facebook started a third-party fact-checking program to rate and review the accuracy of content on their platform in collaboration with IFCN-certified fact-checkers around the globe. Similarly, Google announced Google's News Initiative to fight misinformation, and controversial breaking news. The social media posts contain information in a multi-modal fashion, e.g., a mixture of text, pictures and videos which has evolved the traditional print media to an online based multimedia. The multiple information modality makes the claim more believable and presents a new horizon of opportunities to detect features in fake news[2]. A powerful and competent method for detection of misinformation has recently attracted serious attention from different research communities.

Our major contributions of this paper are summarized as follows:
- We discuss the definition of fake news and the various terms related to it that are used interchangeably.
- We give an overview of existing fake news detection methods on the two broad categories i.e. 1) Content and 2) Social Context of the news.

- We present the publicly available datasets and various approaches which have been adopted to gather the data for this research purpose.
- In addition to this we provide a comprehensive literature survey of the methods used for fake news detection.

The remainder of this paper is organized as follows. In Section 2, we present the definition of fake news, rumour, propaganda and other related terms. In Section 3, we provide an overview of the methods to detect fake news. In Section 4, we summarize the existing datasets. We briefly discuss the existing literature on fake news detection and related areas in Section 5 and finally, we conclude this survey in Section 6.

## 2. FAKE NEWS, RELATED TERMS AND VARIOUS DEBUNKING TOOLS

According to Cambridge Dictionary **Fake News** is defined as "false stories that are created and spread on the Internet to influence public opinion and appear to be true". Fake news is not a new term and has a long legacy reaching back centuries since the development of the earliest writing systems but with the advent of social media the past decade has seen a shift in how the news is propagated that is quite different from the traditional media. The social media platforms have become fertile ground for computational propaganda, and trolling. There are several terms that are used interchangeably for fake news like satire, yellow journalism, hoax, propaganda, misinformation, disinformation, rumor etc, some of them are described below. Figure 1 gives the visual description about the same.

**Propaganda:** Propaganda refers to news stories which are created and propagated by a political entity to influence political view.

**Misinformation:** It is inaccurate information that is deliberately created and is intentionally or unintentionally disseminated disregarding the true intent.

**Disinformation**: It refers to a false or incomplete information that is disseminated with the intention to manipulate facts and mislead the target audience.

**Rumors and hoaxes** are interchangeably used to refer to deliberate falsification or fabrication of information that is constructed to seem valid. They present the unverified and inaccurate claims as validated by traditional news outlets.

**Parody and Satire** usually use humor to give news updates and typically mimic mainstream news media.

**Clickbait**: Sensational headlines are often used as clickbaits to draw the attention of users and encourage them to click and thus redirecting the reader to a different site. More clicks on the advertisements means more money.

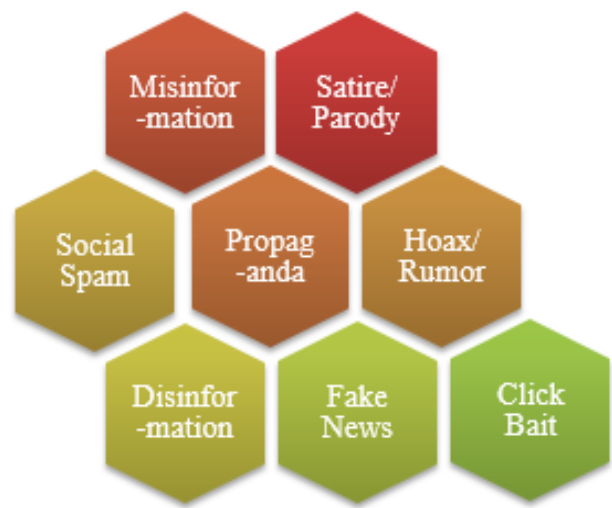

**Figure 1: Key terms related to Fake News**

The rise in the use of propaganda, hoaxes, satire, along with real news and credible content makes it challenging for regular Internet users to distinguish between real and fake news content. But there are various online tools available for debunking fake news like AltNews, APF Fact Check, BSDetector, Hoaxy, Reverse Image Search, Snopes, PolitiFact, Additionally, there are various IFCN-certified fact-checkers around the world that review and rate the credibility of content on various online platforms.

## 3. FAKE NEWS DETECTION METHODS

The wide usage of social media platforms worldwide has provided a fertile ground for the widespread dissemination of online fake news in an unprecedented way. The social network is flooded with massive, diverse, and heterogeneous information (both real and fake), and spreads rapidly on these platforms causing severe impact to the whole society. Therefore, many researchers and technical giants are working together to detect fake news on online media. The traditional automatic rumour detection methods were based on hand crafted feature but with the advent of big data and a huge base of user generated data we have seen a shift to deep-level features. In this section, we discuss various state-of-the-art studies on fake news detection under the broader umbrella of content and social context of the news article.

### 3.1. CONTENT BASED

The content-based fake news detection method aims to detect fake news by analyzing the content[3] of the article , i.e., either the text or image or both within the news article. For automatically detecting the fake news, the researchers often relying on either latent [4],[5],[6],[7],[8],[9] or hand-crafted features [10] of the content.

#### 3.1.1. KNOWLEDGE BASED

Knowledge-based approaches utilize fact checking method in which the given claim is compared with the external sources to verify the authenticity of the given claim. The existing fact checking methods can be categorized as manual (using experts or by crowdsourcing) and automatic fact checking.

**Manual fact checking**
The manual fact-checking can be broadly divided into (I) expert-based and (II) crowd-sourced fact-checking.
**Expert Based:** The expert-based methods use expert-oriented approach and rely on human experts working in specific domains for decision making. The fact-checking websites like Snopes, PoltiFact, GossipCop use this approach. These methods are reliable but are time consuming and do not scale well with the huge volume of content available on social media. Many researches use these websites for creating their own datasets among these are the benchmark datasets LIAR[11] and FakeNewsNet[12].
**Crowd sourced:** For crowdsourced approaches, "wisdom of crowd" helps to check the accuracy of the news articles. A similar approach is used by Fiskkit that provides a platform for people to discuss important news articles and finds out their accuracy. Crowd-sourced fact-checking is even though relatively difficult to manage, biased, has conflicting annotations, is less credible but has better scalability as compared to expert-based fact-checking[13]. CREDBANK[14] is a publicly available large-scale benchmark fake news dataset that is annotated by fact-checkers. The datasets that are created using this approach needs to be filtered for non-credible users and the conflicting annotations need to be resolved beforehand. Some similar datasets can also be created and then be annotated using crowd-sourcing marketplaces such as AMT (Amazon Mechanical Turk).

**Automatic fact checking**
The Manual fact-checking approaches do not scale well with the huge volume of data especially generated with the use of social media as a result, to address this issue automatic fact-checking techniques have been deployed. Instead of relying on human intelligence these methods heavily rely on Natural Language Processing (NLP), Data Mining, Machine Learning (ML) techniques and network/graph theory[13]. The automatic fact-checking process can be divided into two stages: (1) fact extraction which is related to collection of facts and construction of a Knowledge Base and (2) fact-checking which is used to assess the authenticity of news articles by comparing that with the facts in the knowledge base. It uses open web source and knowledge base/graph to check whether the given claim is true/false. The real-world datasets for fake news detection are usually incomplete, unstructured, unlabeled and noisy[15] which make automatic detection a bit complex.

#### 3.1.2. STYLE BASED

Style-based fake news detection follows the same approach like knowledge-based fake news detection of analyzing the news content. However, instead of evaluating the authenticity of news content this method assess the intention of writer to mislead the public[13]. Fake news publishers usually have an intent to influence large communities while spreading distorted and misleading information. To make the titles catchy fake titles use mostly all capitalized words, significantly more proper nouns, and fewer stop words [16]. Style-based approaches captures the distinguishing characteristics of writing styles between legitimate users and anomalous accounts to detect fake news. [17] reports the writing style analysis of hyperpartisan news in connection to fake news. The major contribution of [18] is detecting stylistic deception in written documents.

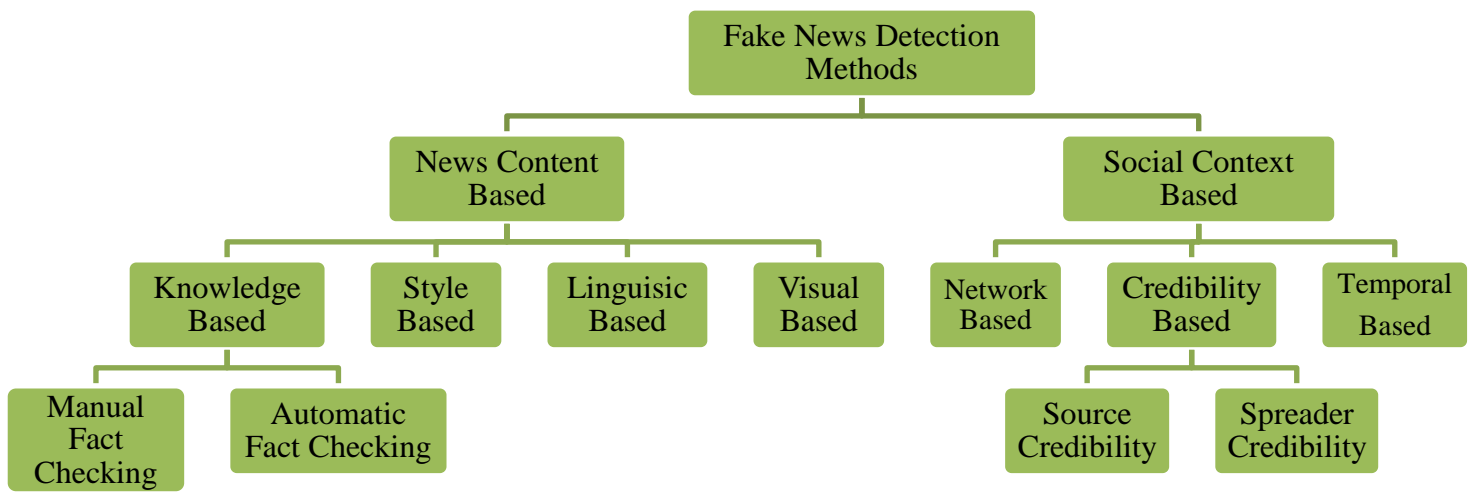

**Figure 2. Fake News Detection Methods**

### 3.1.3. LINGUISTIC BASED

Twenty-six linguistic based textual features were proposed in [19]. In [20] authors proposed an enhanced set of linguistic features to discriminate between fake and real news. [21] used network account features in addition to the linguistic features whereas [22] proposed Social Article Fusion (SAF) model that uses social engagements features along with linguistic. The authors in [23] have used linguistic features along with syntactic and semantic features to distinguish real form a fake news content. [7] proposes model to detect fake news having different lengths of news claims by using different variations of word embedding. [3] investigates a given news content at lexicon-level, syntax-level, semantic-level, and discourse-level. HDSF proposed in [24] learns a hierarchical structure for a given document by dependency parsing at the sentence level. Despite the success of this method in various scenarios, it poses a limitation in case of detecting misinformation on popular social media platforms where the messages are short and thus, the linguistic features extracted from them are often inadequate for machine learning algorithms to make accurate predictions[25]. Additionally, these approaches cannot be used to detect fake news that contains no text content but only a photo or a video.

### 3.1.4. VISUAL BASED

Visual content is often viewed as evidence that can increase the credibility of the news article[2] and hence the fake news publishers tend to utilize provocative visual content to attract and mislead readers. In [10] various visual and statistical image features are extracted for news authentication. Verifying Multimedia Use task [26] under the MediaEval-16 benchmark addresses the problem of detecting digitally manipulated (tampered) images.

### 3.2. SOCIAL CONTEXT BASED

There are three major aspects of the social context i.e.: user profiles, user posts and responses, and network structures[12]. It represents how the news proliferates over time and provides useful information to determine the veracity and stance of news articles. Recent studies [27]–[32] have explored various context based approaches for fake news detection.

### 3.2.1. NETWORK BASED

Network-based fake news detection studies different social networks like friendship, tweet-retweet, post-repost networks to detect fake news. It detects who spreads the fake news, relationships among the spreaders and how fake news propagates on social networks. Users tend to create various networks on online platforms media in terms of their common interests and similarities, these networks serve as paths for information diffusion. [33], [27], [34] studies various networks on social media which gives valuable insights about spreaders of the news and how spreaders connect with each other. [35] models the pattern of message propagation as a tree, which along with the relation among posts give additional information about the temporal behavior and the sentiment of the posts.

### 3.2.2. TEMPORAL BASED

Studies have shown that news stories on the Internet are not static but are constantly evolved over time by adding new information or twisting the actual claim. This is very much evident in cases where the rumors resurge multiple times after the original news article is posted. The lifecycle analysis of rumor helps in understanding this phenomenon and [36] examines the recurring rumors at the message level across different time periods. [37] provides deep understanding into the diffusion patterns of rumors over time.

### 3.2.3. CREDIBILITY BASED

The credibility of claim, publisher, spreader is often assessed by its news quality and trustworthiness/ credibility. [38] identifies the users spreading rumor by leveraging the concept of believability. [39] focusses on assessing credibility of the given claim. [40] proposed a credibility analysis system for evaluating credibility of a Tweet and prevents the proliferation of fake or malicious information. TweetCred is a web-based system that evaluates credibility of the tweet in real time.

## 4. DATASETS FOR FAKE NEWS DETECTION

With several researchers working on this area there are many various fake news datasets that are available, but only a few benchmark fake news datasets are released publicly. In [41] the authors have highlighted the key requirements like homogeneity in length, news genres, topics etc., that are need for creating a reliable fake news detection dataset along with the collection of both real and fake news articles to verify the ground truth for each element in the dataset. For the authors, key factors include. We here present several publicly available datasets along with their comparative analysis. Table 1 gives a brief overview of various fake news datasets available.

**Table 1. Dataset Characteristics**

| Dataset | Task | Task Label | Content Type | Total claims |
|---|---|---|---|---|
| BuzzFace[42] | news veracity assessment and social bot detection | mostly true, mostly false, mixture of true and false, and no factual content. | news items posted to Facebook by nine news outlets | 2263 articles |
| CredBank[43] | Credibility assessment | 5-point credibility scale | news items posted on Twitter | 1049 real-world events |
| Emergent[44] | Rumor stance detection | true, false or unverified | world and national U.S. news and technology stories. | 300 claims and 2,595 associated news articles |

| FacebookHoax[45] | Fake news detection | Hoax, non-hoax | Scientific, conspiracy | 15.5K |
|---|---|---|---|---|
| Fake News Challenge | Stance detection | Agreed, disagrees, discusses, unrelated | News Article | 50K |
| FakeNewsNet[32] | Fake news detection | | US Politics, Entertainment | |
| FakeNewsVsSatire [46] | Fake news detection | fake news or satire | American politics | |
| Fa-Kes [47] | Fake news detection | Fake, True | Syrian War | 804 |
| FEVER[48] | Fact extraction | Supported, refuted, not enough info | | 185K |
| LIAR[11] | Fake news detection | pants-fire, false, barely true, half-true, mostly-true, and true | Political statements | 12.8K |
| MediaEval-VMU [26] | Fake News Detection | Fake, real, unknown | it contains posts related to the 17 events | 15.8K posts |
| PHEME [49] | Rumor Detection and Veracity Classification | True, False or Unverified | nine different newsworthy events | 4,842 tweets |
| RumorEval [50] | Detecting and verifying rumours | supporting, denying, querying or commenting | well-known breaking news | 5.5K Tweets |

## 5. REVIEW OF LITERATURE OF TECHNIQUES USED

Traditionally, the majority of approaches for detecting fake news focus on analyzing the textual content only and utilized hand crafted textual features for the same. But, with an increasing number of articles which are attached with images over the Internet and the extensive use of social media networks, the multimodal features and social-context play a very vital role in better understanding the overall heuristics of the content. The traditional machine learning and rule-based algorithms are inefficient to detect the patterns in today's information age. Hence, to take advantage of big data Deep learning techniques are investigated for fake news detection.

Jin et al. [51] first proposed att-RNN which fuses the multimodal features i.e. textual and visual features along with the user profile features and uses attention mechanism for feature alignment. Wang et al [52] proposed a model named Event Adversarial Neural Network (EANN) that detects fake news on newly arrived events, which can learn both textual and visual features and measures the dissimilarities among different events using an event discriminator. A similar idea is used by Zhang et al. [53] that proposes an end-to-end Multi-modal Knowledge-aware Event Memory Network (MKEMN) that not only learns event invariant features among different events but also exploits the external knowledge connections for accurate news verification. Khattar et al. [8] proposed a Multimodal Variational Autoencoder (MVAE) trained to find correlations across different modalities in a given tweet at post-level. Similarity-Aware Multi-modal Fake News Detection SAFE [54] is proposed by Zhou et al. that jointly learns the textual and visual features in a news article and similarity between them to evaluate whether the news is fake or not. C Song et.al. [55] proposed a Crossmodal Attention Residual and Multichannel convolutional neural Networks (CARMN) which mitigates the influence of noise generated by crossmodal fusion of components at post-level. FND-SCTI proposed by Zeng et.al in [6] not only mines the Semantic Correlations between Text content and Images attached but also uses attention mechanism to highlight the important parts of the given news article. Nguyen et al.[56] proposed FANG, a graph learning based fake news detection framework that captures the social interactions between users, articles, and media for source factuality and fake news prediction. Cui et al. [57] combines the multi-modal data with adversarial learning and incorporates user sentiment with it. An end-to-end neural network based model that automatically calculates the credibility of the given claims was proposed by Popat et al. [58]. Similarly to this, FAKEDETECTOR proposed by Zhang et al. [59] aims to provide the credibility labels to not only the claim, but also of the creators and subjects simultaneously.

Table 2. Literature Review

| Features / Model | Content based | | Social Context based | | |
|---|---|---|---|---|---|
| | Text Based | Visual Based | Credibility Features | Temporal Based | Network Based |
| Att-RNN[51] | ✓ | ✓ | | ✓ | ✓ |
| DeClarE[58] | ✓ | | ✓ | | |
| EANN[52] | ✓ | ✓ | | | |
| MKEMN[53] | ✓ | ✓ | | ✓ | ✓ |
| MVAE[8] | ✓ | ✓ | | | |
| SAFE[54] | ✓ | ✓ | | | |
| CARMN[55] | ✓ | ✓ | | | |
| FAKEDETECTOR[59] | ✓ | | ✓ | | |
| FANG[56] | ✓ | ✓ | ✓ | ✓ | ✓ |
| FND-SCTI[6] | ✓ | ✓ | | | |
| SAME[57] | ✓ | ✓ | ✓ | | |

## 6. CONCLUSION

With an increase in the popularity and usage of social media over the past few years, a huge population of readers prefer to consume news from social media instead of traditional news media. Keeping this in mind, many publishers use social media and Internet in general as a breeding grounds for spreading propaganda and rumours rapidly which has strong negative impacts on the society. In this text we have mentioned several freely available Fake News Detection tools that should be used so that we forward only credible and genuine news. In this paper, we have explored the present fake news detection methods by reviewing existing literature under two categories: The Content Based and The Social Context Based Fake news detection. In the content based method, the article/post is considered that may contain the textual or visual content or both. In the social context based method, the propagation structure and the credibility of the publisher is considered. While the content based methods can be used for early detection of fake news the context based methods fail to do so because of the absence of the propagation details in the very beginning of the proliferation of misinformation. Additionally, despite many researchers are focusing on this area but still there are only a few publicly available benchmark datasets.

**REFERENCES**

[1] X. Zhang and A. A. Ghorbani, "An overview of online fake news: Characterization, detection, and discussion," *Inf. Process. Manag.*, vol. 57, no. 2, pp. 1–26, 2020, doi: 10.1016/j.ipm.2019.03.004.
[2] J. Cao, P. Qi, Q. Sheng, T. Yang, J. Guo, and J. Li, "Exploring the role of visual content in fake news detection," *arXiv*, no. April, 2020, doi: 10.1007/978-3-030-42699-6_8.
[3] X. ZHOU, A. JAIN, V. V. PHOHA, and R. ZAFARANI, "Fake news early detection: A theory-driven model," *arXiv*, vol. 1, no. 2, pp. 1–25, 2019.
[4] Y. Wang *et al.*, "EANN in FND-2018.pdf," pp. 849–857, 2018.
[5] X. Z. B, J. Wu, and R. Zafarani, *SAFE : Similarity-Aware Multi-modal Fake*. Springer International

Publishing, 2020.

[6] J. Zeng, Y. Zhang, and X. Ma, "Fake news detection for epidemic emergencies via deep correlations between text and images," *Sustain. Cities Soc.*, vol. 66, p. 102652, 2021, doi: 10.1016/j.scs.2020.102652.

[7] M. H. Goldani, S. Momtazi, and R. Safabakhsh, "Detecting fake news with capsule neural networks," *Appl. Soft Comput.*, vol. 101, p. 106991, 2021, doi: 10.1016/j.asoc.2020.106991.

[8] D. Khattar, M. Gupta, J. S. Goud, and V. Varma, "MvaE: Multimodal variational autoencoder for fake news detection," *Web Conf. 2019 - Proc. World Wide Web Conf. WWW 2019*, no. May, pp. 2915–2921, 2019, doi: 10.1145/3308558.3313552.

[9] S. Yoon *et al.*, "Detecting incongruity between news headline and body text via a deep hierarchical encoder," *33rd AAAI Conf. Artif. Intell. AAAI 2019, 31st Innov. Appl. Artif. Intell. Conf. IAAI 2019 9th AAAI Symp. Educ. Adv. Artif. Intell. EAAI 2019*, pp. 791–800, 2019, doi: 10.1609/aaai.v33i01.3301791.

[10] Z. Jin, J. Cao, Y. Zhang, J. Zhou, and Q. Tian, "Novel Visual and Statistical Image Features for Microblogs News Verification," *IEEE Trans. Multimed.*, vol. 19, no. 3, pp. 598–608, 2017, doi: 10.1109/TMM.2016.2617078.

[11] W. Y. Wang, "'Liar, liar pants on fire': A new benchmark dataset for fake news detection," *ACL 2017 - 55th Annu. Meet. Assoc. Comput. Linguist. Proc. Conf. (Long Pap.*, vol. 2, pp. 422–426, 2017, doi: 10.18653/v1/P17-2067.

[12] K. Shu, D. Mahudeswaran, S. Wang, D. Lee, and H. Liu, "FakeNewsNet : A Data Repository with News Content , Social Context and Spatiotemporal Information for Studying Fake News on Social Media," 2018.

[13] X. Zhou and R. Zafarani, "A Survey of Fake News: Fundamental Theories, Detection Methods, and Opportunities," *ACM Comput. Surv.*, vol. 53, no. 5, 2020, doi: 10.1145/3395046.

[14] T. Mitra and E. Gilbert, "CREDBANK: A large-scale social media corpus with associated credibility annotations," *Proc. 9th Int. Conf. Web Soc. Media, ICWSM 2015*, pp. 258–267, 2015.

[15] K. Shu, A. Sliva, S. Wang, J. Tang, and H. Liu, "Fake news detection on social media: A data mining perspective," *arXiv*, vol. 19, no. 1, pp. 22–36, 2017, doi: 10.1145/3137597.3137600.

[16] Benjamin D. Horne and Sibel Adalı, "This Just In: Fake News Packs a Lot in Title, Uses Simpler, Repetitive Content in Text Body, More Similar to Satire than Real News," in *Proceedings of the First Workshop on Fact Extraction and VERification (FEVER)*, 2018, pp. 40–49, doi: arXiv:1703.09398.

[17] M. Potthast, J. Kiesel, K. Reinartz, J. Bevendorff, and B. Stein, "A stylometric inquiry into hyperpartisan and fake news," *ACL 2018 - 56th Annu. Meet. Assoc. Comput. Linguist. Proc. Conf. (Long Pap.*, vol. 1, pp. 231–240, 2018, doi: 10.18653/v1/p18-1022.

[18] S. Afroz, M. Brennan, and R. Greenstadt, "Detecting hoaxes, frauds, and deception in writing style online," *Proc. - IEEE Symp. Secur. Priv.*, pp. 461–475, 2012, doi: 10.1109/SP.2012.34.

[19] S. Hakak, M. Alazab, S. Khan, T. R. Gadekallu, P. K. R. Maddikunta, and W. Z. Khan, "An ensemble machine learning approach through effective feature extraction to classify fake news," *Futur. Gener. Comput. Syst.*, vol. 117, pp. 47–58, 2021, doi: 10.1016/j.future.2020.11.022.

[20] G. Gravanis, A. Vakali, K. Diamantaras, and P. Karadais, "Behind the cues: A benchmarking study for fake news detection," *Expert Syst. Appl.*, vol. 128, pp. 201–213, 2019, doi: 10.1016/j.eswa.2019.03.036.

[21] D. Mouratidis, M. N. Nikiforos, and K. L. Kermanidis, "Deep learning for fake news detection in a pairwise textual input schema," *Computation*, vol. 9, no. 2, pp. 1–15, 2021, doi: 10.3390/computation9020020.

[22] K. Shu, D. Mahudeswaran, and H. Liu, "FakeNewsTracker: a tool for fake news collection, detection, and visualization," *Comput. Math. Organ. Theory*, vol. 25, no. 1, pp. 60–71, 2019, doi: 10.1007/s10588-018-09280-3.

[23] and R. M. Verónica Pérez-Rosas, Bennett Kleinberg, Alexandra Lefevre, "Automatic detection of fake news," *CEUR Workshop Proc.*, vol. 2789, pp. 168–179, 2020.

[24] H. Karimi and J. Tang, "Learning hierarchical discourse-level structure for fake news detection," *NAACL HLT 2019 - 2019 Conf. North Am. Chapter Assoc. Comput. Linguist. Hum. Lang. Technol. - Proc. Conf.*, vol. 1, pp. 3432–3442, 2019, doi: 10.18653/v1/n19-1347.

[25] Y. Liu and Y. B. Wu, "Early Detection of Fake News on Social Media Through Propagation Path

Classification with Recurrent and Convolutional Networks," pp. 354–361.
[26] C. Boididou *et al.*, "Verifying Multimedia Use at MediaEval 2016," *CEUR Workshop Proc.*, vol. 1739, pp. 4–6, 2016.
[27] K. Shu, S. Wang, and H. Liu, "Beyond news contents: The role of social context for fake news detection," *arXiv*, no. i, pp. 312–320, 2017.
[28] K. Shu, S. Wang, and H. Liu, "Understanding User Profiles on Social Media for Fake News Detection," *Proc. - IEEE 1st Conf. Multimed. Inf. Process. Retrieval, MIPR 2018*, pp. 430–435, 2018, doi: 10.1109/MIPR.2018.00092.
[29] K. Shu, H. Russell Bernard, and H. Liu, "Studying fake news via network analysis: Detection and mitigation," *arXiv*, pp. 836–837, 2018, doi: 10.1007/978-3-319-94105-9_3.
[30] K. Shu, H. Russell Bernard, and H. Liu, "Studying fake news via network analysis: Detection and mitigation," *arXiv*, 2018, doi: 10.1007/978-3-319-94105-9_3.
[31] A. Bodaghi and J. Oliveira, "The characteristics of rumor spreaders on Twitter: A quantitative analysis on real data," *Comput. Commun.*, vol. 160, pp. 674–687, 2020, doi: 10.1016/j.comcom.2020.07.017.
[32] K. Shu, D. Mahudeswaran, S. Wang, D. Lee, and H. Liu, "FakeNewsNet: A Data Repository with News Content, Social Context, and Spatiotemporal Information for Studying Fake News on Social Media," *Big Data*, vol. 8, no. 3, pp. 171–188, 2020, doi: 10.1089/big.2020.0062.
[33] X. Zhou and R. Zafarani, "Network-based Fake News Detection: A Pattern-driven Approach," *arXiv*, 2019, doi: 10.1145/3373464.3373473.
[34] N. Ruchansky, S. Seo, and Y. Liu, "CSI: A hybrid deep model for fake news detection," *Int. Conf. Inf. Knowl. Manag. Proc.*, vol. Part F1318, pp. 797–806, 2017, doi: 10.1145/3132847.3132877.
[35] K. Wu, S. Yang, and K. Q. Zhu, "False Rumors Detection on Sina Weibo by Propagation Structures," pp. 651–662, 2015.
[36] J. Shin, L. Jian, K. Driscoll, and F. Bar, "The diffusion of misinformation on social media: Temporal pattern, message, and source," 2018, doi: 10.1016/j.chb.2018.02.008.
[37] S. Kwon, M. Cha, and K. Jung, "Rumor detection over varying time windows," *PLoS One*, vol. 12, no. 1, pp. 1–19, 2017, doi: 10.1371/journal.pone.0168344.
[38] B. Rath, W. Gao, J. Ma, and J. Srivastava, "From Retweet to Believability: Utilizing Trust to Identify Rumor Spreaders on Twitter," *Soc. Netw. Anal. Min.*, vol. 8, no. 1, pp. 179–186, 2018, doi: 10.1007/s13278-018-0540-z.
[39] N. Sitaula, C. K. Mohan, J. Grygiel, X. Zhou, and R. Zafarani, "Credibility-based Fake News Detection," *arXiv*, 2019, doi: 10.1007/978-3-030-42699-6_9.
[40] M. Alrubaian, M. Al-Qurishi, M. M. Hassan, and A. Alamri, "A Credibility Analysis System for Assessing Information on Twitter," *IEEE Trans. Dependable Secur. Comput.*, vol. 15, no. 4, pp. 661–674, 2018, doi: 10.1109/TDSC.2016.2602338.
[41] V. L. Rubin, Y. Chen, and N. J. Conroy, "Deception detection for news: Three types of fakes," *Proc. Assoc. Inf. Sci. Technol.*, vol. 52, no. 1, pp. 1–4, 2015, doi: 10.1002/pra2.2015.145052010083.
[42] G. C. Santia and J. R. Williams, "BuzzFace : A News Veracity Dataset with Facebook User Commentary and Egos," no. Icwsm, pp. 531–540, 2018.
[43] T. Mitra and E. Gilbert, "CREDBANK : A Large-Scale Social Media Corpus with Associated Credibility Annotations," pp. 258–267.
[44] W. Ferreira and A. Vlachos, "Emergent : a novel data-set for stance classification," no. 1, pp. 1163–1168, 2016.
[45] E. Tacchini, G. Ballarin, M. L. Della Vedova, S. Moret, and L. De Alfaro, "Some Like it Hoax : Automated Fake News Detection in Social Networks," pp. 1–12.
[46] J. Golbeck *et al.*, "Fake news vs satire: A dataset and analysis," *WebSci 2018 - Proc. 10th ACM Conf. Web Sci.*, no. May, pp. 17–21, 2018, doi: 10.1145/3201064.3201100.
[47] F. K. A. Salem, R. Al Feel, S. Elbassuoni, M. Jaber, and M. Farah, "FA-KES : A Fake News Dataset around the Syrian War," no. Icwsm, 2019.
[48] J. Thorne, A. Vlachos, C. Christodoulopoulos, and A. Mittal, "FEVER : a large-scale dataset for Fact Extraction and VERification," pp. 809–819, 2018.
[49] A. Zubiaga *et al.*, "Analysing How People Orient to and Spread Rumours in Social Media by Looking at Conversational Threads," pp. 1–34.

[50] G. Gorrell, K. Bontcheva, L. Derczynski, E. Kochkina, M. Liakata, and A. Zubiaga, "RumourEval 2019: Determining rumour veracity and support for rumours," *arXiv*, 2018.

[51] Z. Jin, J. Cao, H. Guo, Y. Zhang, and J. Luo, "Multimodal fusion with recurrent neural networks for rumor detection on microblogs," *MM 2017 - Proc. 2017 ACM Multimed. Conf.*, pp. 795–816, 2017, doi: 10.1145/3123266.3123454.

[52] Y. Wang *et al.*, "EANN: Event adversarial neural networks for multi-modal fake news detection," *Proc. ACM SIGKDD Int. Conf. Knowl. Discov. Data Min.*, pp. 849–857, 2018, doi: 10.1145/3219819.3219903.

[53] H. Zhang, Q. Fang, S. Qian, and C. Xu, "Multi-modal knowledge-aware event memory network for social media rumor detection," *MM 2019 - Proc. 27th ACM Int. Conf. Multimed.*, pp. 1942–1951, 2019, doi: 10.1145/3343031.3350850.

[54] X. Z. B, J. Wu, and R. Zafarani, "SAFE : Similarity-Aware Multi-modal Fake," pp. 354–367, 2020, doi: 10.1007/978-3-030-47436-2.

[55] C. Song, N. Ning, Y. Zhang, and B. Wu, "A multimodal fake news detection model based on crossmodal attention residual and multichannel convolutional neural networks," *Inf. Process. Manag.*, vol. 58, no. 1, p. 102437, 2021, doi: 10.1016/j.ipm.2020.102437.

[56] V. H. Nguyen, K. Sugiyama, P. Nakov, and M. Y. Kan, "FANG: Leveraging Social Context for Fake News Detection Using Graph Representation," *Int. Conf. Inf. Knowl. Manag. Proc.*, pp. 1165–1174, 2020, doi: 10.1145/3340531.3412046.

[57] L. Cui, S. Wang, and D. Lee, "Same: Sentiment-aware multi-modal embedding for detecting fake news," *Proc. 2019 IEEE/ACM Int. Conf. Adv. Soc. Networks Anal. Mining, ASONAM 2019*, pp. 41–48, 2019, doi: 10.1145/3341161.3342894.

[58] K. Popat, S. Mukherjee, A. Yates, and G. Weikum, "Declare: Debunking fake news and false claims using evidence-aware deep learning," *Proc. 2018 Conf. Empir. Methods Nat. Lang. Process. EMNLP 2018*, pp. 22–32, 2020, doi: 10.18653/v1/d18-1003.

[59] J. Zhang, B. Dong, and P. S. Yu, "FakeDetector: Effective fake news detection with deep diffusive neural network," *Proc. - Int. Conf. Data Eng.*, vol. 2020-April, pp. 1826–1829, 2020, doi: 10.1109/ICDE48307.2020.00180.